\documentclass[letterpaper]{article}
\usepackage{natbib,alifeconf}
\usepackage{topcapt}
\usepackage{booktabs}

\usepackage[colorinlistoftodos]{todonotes}

\newcommand{\be}{\begin{eqnarray}}
\newcommand{\ee}{\end{eqnarray}}

\title{Evolution leads to a diversity of motion-detection neuronal circuits}
\author{Ali Tehrani-Saleh$^{1,2}$, Thomas LaBar$^{2,3,4}$ \and Christoph Adami$^{2,3,4,5,\dagger}$ \\
\mbox{}\\
$^1$Department of Computer Science and Engineering \\
$^2$BEACON Center for the Study of Evolution in Action \\
$^3$Department of Microbiology and Molecular Genetics \\
$^4$Program in Ecology, Evolutionary Biology, and Behavior \\
$^5$Department of Physics and Astronomy, Michigan State University, East Lansing, MI 48824\\
$^\dagger$adami@msu.edu}

\begin{document}
\maketitle

\begin{abstract}
A central goal of evolutionary biology is to explain the origins and distribution of diversity across life. Beyond species or genetic diversity, we also observe diversity in the circuits (genetic or otherwise) underlying complex functional traits. However, while the theory behind the origins and maintenance of genetic and species diversity has been studied for decades, theory concerning the origin of diverse functional circuits is still in its infancy. It is not known how many different circuit structures can implement any given function, which evolutionary factors lead to different circuits, and whether the evolution of a particular circuit was due to adaptive or non-adaptive processes. Here, we use digital experimental evolution to study the diversity of neural circuits that encode motion detection in digital (artificial) brains. We find that evolution leads to an enormous diversity of potential neural architectures encoding motion detection circuits, even for circuits encoding the exact same function. Evolved circuits vary in both redundancy and complexity (as previously found in genetic circuits) suggesting that similar evolutionary principles underlie circuit formation using any substrate. We also show that a simple (designed) motion detection circuit that is optimally-adapted gains in complexity when evolved further, and that selection for mutational robustness led this gain in complexity.

\end{abstract}

\section{Introduction}

One of the most astonishing aspects of life is the overwhelming amount of diversity that has existed throughout life's history. Ever since Charles Darwin published {\em On the Origin of Species}, evolutionary biologists have tried to understand the processes that lead to biological diversity~\citep{darwin2006origin}. On the micro scale, the question of how genetic diversity is maintained within a population has been of interest to population geneticists~\citep{kimura1964number,lewontin1966molecular,sved1967number,ayala1974frequency} for decades; work on this topic still continues to this day~\citep{good2017dynamics}. In a similar fashion, ecologists have long been interested in the ecological and evolutionary processes that lead to the origins~\citep{rainey2000emergence,nosil2012ecological} and maintenance~\citep{rosenzweig1995species,chesson2000mechanisms} of species diversity. The rise of cheap sequencing technologies in recent years has led to the recognition of another characteristic of biological diversity, molecular diversity~\citep{tenaillon2012molecular}, or diversity in the sense that multiple genotypes can lead to the same phenotype~\citep{gonzalez2017adaptive}. In other words, evolution can lead to a diversity of genetic circuits across species~\citep{tsong2003evolution}.

The evolutionary principles that lead to molecular diversity in genetic systems has been well-explored. The relationship between genotype and phenotype must be many-to-one to allow for the existence of neutral evolutionary trajectories between genotypes. Computational studies of metabolic networks, gene regulatory networks, and RNA-structure networks [reviewed in~\citep{wagner2011origins}] all show evidence of neutral paths that conserve phenotypes between different genotypes. Many-to-one genotype-phenotype mappings are even present in artificial digital evolution systems [e.g.,~\citep{labar2016evolvability,fortuna2017genotype,labar2017evolution}] and evolutionary simulations of digital logic circuits~\citep{raman2010evolvability}. Empirical studies of bio\-logical systems suggest the existence of multiple genotypes encoding similar phenotypes, either through genetic analysis~\citep{tsong2006evolution,taylor2016diverse}, comparative genomics~\citep{cross2011evolution}, or experimental evolution~\citep{lind2015experimental,hope2017experimental}. However, the evolutionary reasons why populations evolve one genotype instead of another genotype, or which evolutionary processes lead to the evolution of different genotypes, are largely unexplored in biological systems due to the difficulty of deciphering every possible evolutionary trajectory and process, and the waiting time required for many of these evolutionary events to occur [but see ~\citep{lind2015experimental}]. This difficulty presents a prime opportunity for artificial life and digital evolution studies to perform ``digital genetics" and test hypotheses for why some populations, but not others, evolve certain genotypic characteristics~\citep{adami2006digital}.

Genetic circuits are not the only biological network shaped by evolution. Neuronal circuits are also shaped by selective pressures, and much work has been devoted to understand those. Much of the literature, however, has focused on whether evolution optimizes the wiring patterns of a brain, or the efficiency of the circuitry [see, e.g., the discussion in chapter 7 of~\citep{Sporns2011}]. For example, it is clear that the wiring pattern of the neuronal circuitry of the roundworm {\it Caenorhabditis elegans} is not optimal~\citep{Ahnetal2006}. At the same time, there appear to be certain network motifs that are strongly favored in the worm brain~\citep{Qianetal2011}, suggesting that evolution has a hand in optimizing computational efficiency. However, very little is known about the wiring diversity underlying circuits with the {\em same} function. According to the principles of evolvability and robustness discussed above, such diversity could be key
for the adaptability of brains. In fact, both modeling~\citep{prinz2004similar} and empirical~\citep{goaillard2009functional} studies have shown that neuronal circuits can vary in their internal parameters but lead to the same functional output~\citep{marder2011variability}. And while many of these studies examine variation within one species~\citep{goaillard2009functional}, similar results have also been found between species, suggesting evolutionary mechanisms can also cause these differences~\citep{shomrat2011alternative}. This outcome is not surprising, as evolution and natural selection is expected to primarily act on the function, not the circuit encoding said function~\citep{shomrat2011alternative}. These results motivate the question as to how and why evolution leads to neuronal circuits with different characteristics for the same function.

\begin{figure}[htb]
\centering
\includegraphics[scale=0.15]{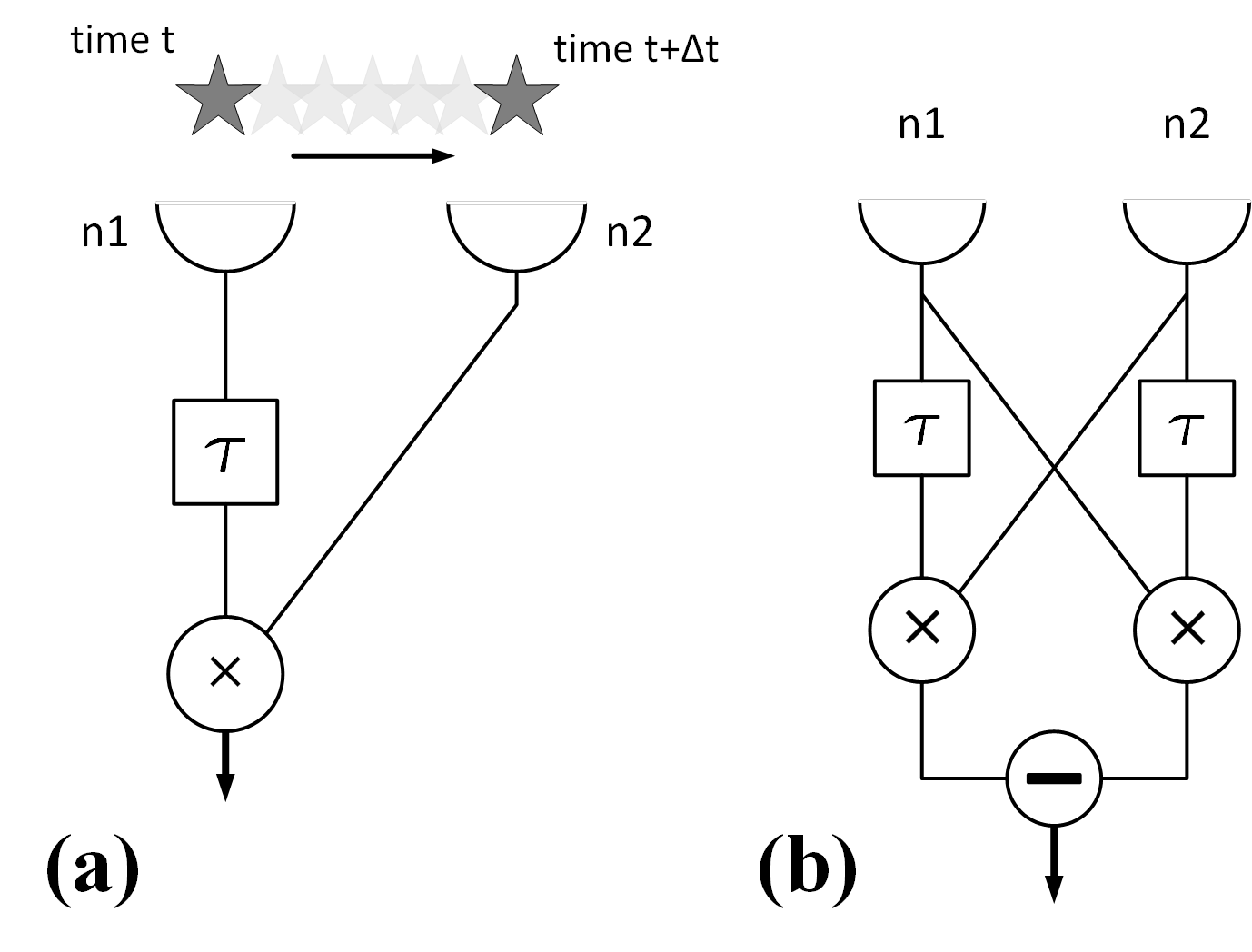}
\caption{a) A half Reichardt detector circuit. An object (star) moving from left to right stimulating two adjacent receptors, n1 and n2, at time points $t$ and $t+\Delta t$. b) A full Reichardt detector circuit. In full Reichardt detector circuits, the results of the multiplications from each half circuit are subtracted.}
\label{RD}
\end{figure}

Here we use digital evolution to study the evolution of neuronal circuits for visual motion detection. Perception of moving objects in the environment is of utmost significance from an evolutionary standpoint since it can be critical to survival of animals (including humans); detecting predators, prey, or falling objects can pose a {\em live or die} question~\citep{palmer1999vision}. In the 1950s, Werner Reichardt along with Bernhard Hassenstein proposed a simple computational model [now known as the Reichardt detector], that is based on a delay-and-compare scheme~\citep{hassenstein1956systemtheoretische}. The main idea behind this model is that a moving object stimulates two adjacent receptors (or regions) in the retina at two different time points. In Fig.~\ref{RD}a, an object (a star) is moving from left to right stimulating two adjacent receptors n1 and n2, at time points $t$ and $t+\Delta t$. In the neural circuit illustrated in Fig.~\ref{RD}a, which is a portion of the entire Reichardt detector circuit, $\tau$ functions as a temporal filter that {\em delays} the received stimulus from receptor n1. This delayed signal will then be multiplied (in the $\times$ neuron) with the stimulus received in n2 at $t+\Delta t$. This multiplication result, therefore, detects motion from left to right. However, this half-circuit only detects motion in one direction. In the full Reichardt detector circuit shown in Fig.~\ref{RD}b, the outcome of the multiplication from two similar computations, but in opposite directions, are subtracted. Thus, the result will be a positive value for left to right motion (also called {\em preferred direction}, PD), and negative for right to left motion, (termed the {\em null direction}, ND).
%Fig. 1

Beyond the Reichardt detector, other types of motion detection models were also proposed, e.g. edge-based models ~\citep{marr1981directional} and spatial-frequency-based models~\citep{adelson1985spatiotemporal}. However, most computational motion detection models are based on the delay-and-compare scheme~\citep{palmer1999vision}. For example, the Barlow-Levick (BL) motion detection model~\citep{barlow1965mechanism} is similar to the Reichardt model in that it also employs asymmetric temporal filtering of signals that are then fed to a non-linearity component, but they differ in the location of the filter and type of non-linearity component. While motion detection in mammals and in particular humans is expected to be far more complex, there are significant similarities to the basic Reichardt detector logic~\citep{borst2015common}, and thus the Reichardt detector ``module" of motion detection is likely a key component of all motion detection circuits. 

Using digital experimental evolution methods, we found that motion detection circuits can be encoded by a wide diversity of neuronal architectures. Evolved brains differ in the logic gates used to perform motion detection, in the wiring between these logic gates, in the presence of redundant logic gates, and in their total complexity (i.e., number of logic gates). We explored the evolutionary significance in complexity variation between brains by evolving brains using a handwritten optimal motion detection circuit as the ancestor. These brains also increased in complexity although no improvement in the performance of their circuit could occur. Instead, these brains evolved greater complexity due to selection for mutational robustness. These results suggest that different species may evolve different circuits for similar neuronal functions. 

\section{Methods}
In this study, we use an agent-based model to study evolution of computational visual motion detection circuits. In this model, agents embody neural networks known as ``Markov brains" (MB)~\citep{hintze2017markov}. Markov brains have three different types of neurons that help the agent interact with the outside world: 1) sensory neurons, that receive the information from the environment, 2) hidden neurons that assimilate the agent's processing unit, and 3) decision (``motor") neurons that function as the actuators of the agent. In other words, sensory neurons are written to by the surrounding environment, hidden neurons process the received information, and the decision neurons specify the actions of the agent in its environment.

\begin{figure}[htb]
\begin{center}
\includegraphics[scale=0.3]{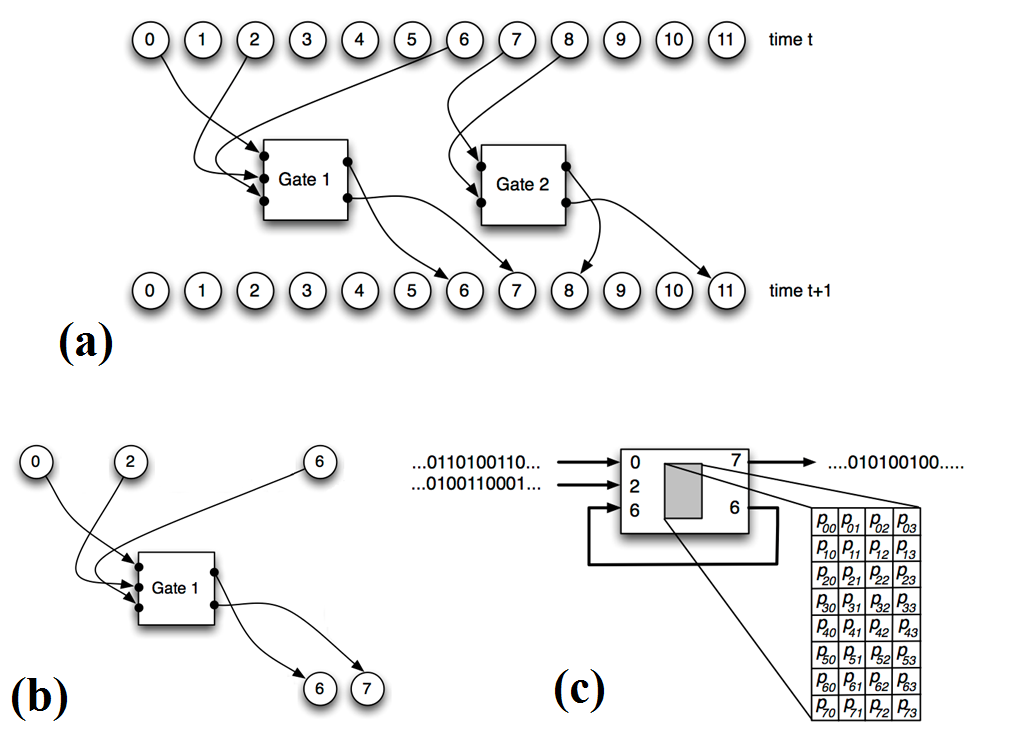}
\caption{a) A Markov brain with 11 neurons and 2 gates shown at two time steps $t$ and $t+1$. The states of neurons at time t and the logic operations of gates determine the states of neurons at time $t+1$. b) One of the gates of the MB whose inputs are neurons 0, 2, and 6 and its outputs are neurons 6 and 7. c) Probabilistic logic table of gate 1.}
\label{mb}
\end{center}
\end{figure}

Markov brains are evolvable networks of neurons in which the neurons are connected via probabilistic/deterministic logic gates. In the experimental setup used in this study, a logic neuron is a binary variable whose state is either 0 or 1 (it is quiescent or it fires\footnote{These logic neurons are thought to represent the state of groups of biological neurons.}). The states of the neurons are updated in a Markov fashion, i.e., the probability distribution of states of the neurons at time step $t+1$ depends only on the states of neurons at time step $t$ as shown in Fig.~\ref{mb}a. That figure shows a Markov brain with 11 neurons and two hidden Markov gates (HMG) at two consecutive time steps $t$ and $t+1$. Hidden Markov gates determine how the states of the neurons at time step t+1 are updated given the states of the neurons at time $t$. For example in Fig.~\ref{mb}b, gate 1 takes the states of neurons 0, 2, and 6 as inputs and writes updated states to output neurons 6 and 7. Each hidden Markov gate has a probabilistic logic table that specifies the probability of every possible output given the states of the input (Fig.~\ref{mb}c). That figure shows the probability table of gate 1 with 8 rows for all possible input states, and 4 columns for each possible output states (note that there are $2^3=8$ possible input states for 3 binary inputs, and similarly, $2^2=4$ for outputs). Each entry in the table represents the probability of a specific output, given a particular input. For instance, $p_{53}$ represents the probability of getting output states $\langle 1,1 \rangle$, with decimal representation 3, given the input states $\langle 1,0,1 \rangle$, with decimal representation 5. As a result, the sum of the probabilities of each row should be equal to 1. In this work, we constrain hidden Markov gates to be deterministic, therefore, the output states will always be the same given a particular input (probabilities in the table are either 0 or 1 and only one entry in each row of the table can be equal to 1).
%Fig. 2
Markov brains can evolve to perform a variety of tasks such as active categorical perception~\citep{blockCatching}, swarming in predator-prey interactions~\citep{confusionIsSufficient}, collision avoidance strategies using optical flow classification in fruit flies~\citep{Tehranietal2016}, and decision making strategies in humans~\citep{kvam2015computational}. In the evolutionary process, the connections of the networks and the underlying logic of the connected gates change (evolve), and therefore, the agents adapt to their environment. More specifically, the number of gates, how each gate is connected to its inputs/outputs neurons, and the logic table of the gates are subject to evolution. However, the total number of neurons, the number of each type of neurons (i.e., sensory neurons, hidden neurons, and decision neurons), does not change during evolution. In our experimental setup for instance, we use MBs with 16 neurons in which two neurons (neurons 1 and 2) are designated as sensory neurons, and two neurons (neurons 15 and 16) are assigned as decision neurons, while the remaining 12 neurons are hidden neurons.
\begin{figure}[htb]
\begin{center}
\includegraphics[scale=0.4]{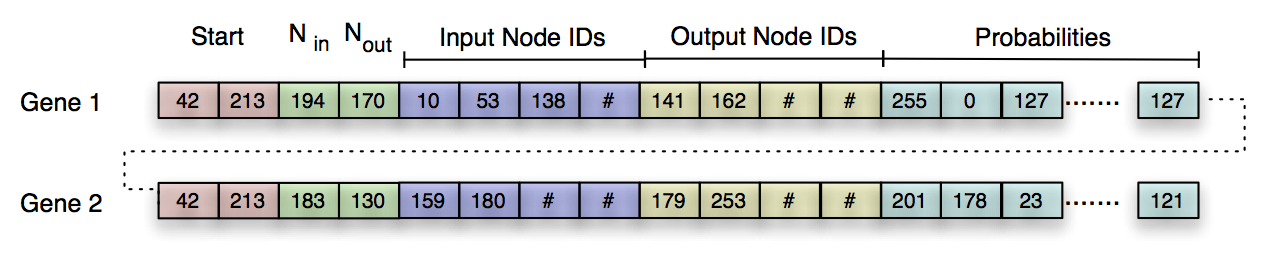}
\caption{A Markov brain is encoded in a sequence of bytes that serves as the agent's genome.}
\label{mb-encoding}
\end{center}
\end{figure}
%Fig. 3
In order to evolve MBs, we apply a Genetic Algorithm (GA) to a population of MBs in which each MB is encoded in a genome as shown in Fig.~\ref{mb-encoding}. The genome of each MB is a sequence of numbers in the range [0,255] (a sequence of bytes) that encodes hidden Markov gates (HMGs), their connections, and their logic. The arbitrary pair $\langle 42,213 \rangle$ is chosen as the start codon for each gate. The next two bytes following the start codon encode the number of inputs and the number of outputs of the HMG, respectively. In our experimental setup, we constrained MBs to always have 2 inputs and 1 output, therefore, these two bytes are ignored in transcription. The subsequent (downstream) loci in the genome encode which neurons are connected to this HMG as input, which neuron is connected to the output, and finally the logic table of the HMG.

In our experimental setup, we initialized the populations with 100 genomes with 5,000 random bytes. We sprinkled those random bytes with four start codons in each genome to speed up initial evolution. Thus, all genomes in the initial population have at least four random HMGs. As mentioned before, all HMGs in our setup are deterministic and have 2 inputs and 1 output. As a result, HMGs can only have 16 possible logic tables. We ran 100 replicates of this experiment for 10,000 generations with mutations, roulette wheel selection, and 5\% elitism. The GA configuration is presented in more detail in Table~\ref{GAtable}. 

\begin{table}[!hbt]
\center{
\caption{Genetic Algorithm configuration. We evolved 100 populations of 100 MBs for 10,000 generations with point mutations, deletions, and insertion. We used roulette wheel selection, with 5\% elitism, and with no cross-over or immigration.}
\label{GAtable}
\begin{tabular}{|c|c|}\hline
Population size & 100 \\ \hline
Generations & 2000 \\ \hline
Initial genome length & 5,000 \\ \hline
Initial start codons & 4 \\ \hline
Point mutation rate & 0.5\% \\ \hline
Gene deletion rate & 2\%  \\ \hline
Gene duplication rate  & 5\% \\ \hline
Elitism & 5\% \\ \hline
Crossover & None \\ \hline
Immigration & None \\ \hline
\end{tabular}
}
\vskip 0.25cm
\end{table}

\begin{figure}[htb]
\centering
\includegraphics[scale=0.22]{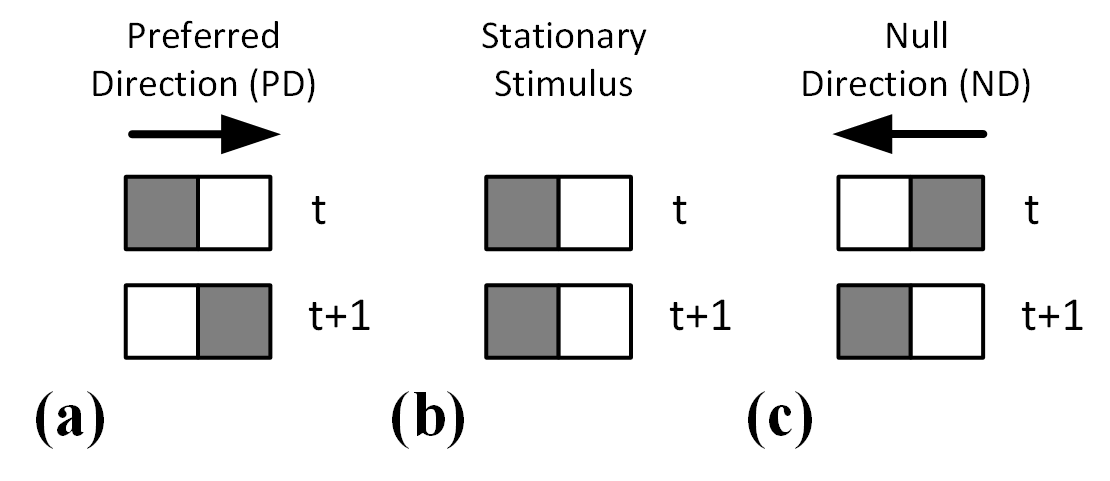}
\caption{Schematic examples of three types of input patterns received by the two sensory neurons at two consecutive time steps. Grey squares show presence of the stimuli in those neurons. a) Preferred direction (PD). b) Stationary stimulus. c) Null direction (ND).}
\label{fig4}
\end{figure}

The fitness function is designed in order to evolve MBs that function as a visual motion detection circuit. In doing so, two sets of stimuli are presented to the agent in two consecutive time steps and the agent classifies the input as either: motion in preferred direction (PD), stationary, or motion in null direction (ND). Neurons 1 and 2 (the sensory neurons) represent two adjacent receptors separated by a fixed distance that can sense the presence or the absence of a visual stimulus. The binary value of the sensory neuron becomes 1 when a stimulus is present, and it becomes (or remains) 0 otherwise (see Fig.~\ref{fig4}). Thus, there are 16 possible sensory patterns that can be presented to the agent (2 binary neurons at 2 time steps). Among these 16 input patterns, 3 input patterns are PD, 3 are ND, and the other 10 are stationary patterns. Agents classify the sensory pattern with 2 decision neurons, neurons 15 and 16. We assigned the sum of the values of the decision neurons to represent the category of the sensory pattern: when both decision neurons fire (sum=2), the sensory pattern is classified as PD, when only one of the decision neurons fires (sum=1), the sensory pattern is classified as stationary, and when neither fire the sensory pattern is classified as ND (sum=0). We chose this encoding for three classes of input pattern to facilitate the evolution of motion detection circuits. In preliminary experiments, we tried three different methods of encoding input pattern classes and found this one to evolve the fastest. In those preliminary experiments, we tried  the following alternative encodings: assigning one neuron to each class (i.e. three decision neurons), assigning the decimal value of the pair of decision neurons to each class, i.e., $00 \rightarrow$ ND, $01 \rightarrow$ stationary, $10 \rightarrow$ PD, and ignore 11, and finally assigning the sum of the values of decision neurons to each class. For the last two encoding methods, we tried all possible permutations of encodings and the one we chose consistently leads to the best results.
%Fig. 4

All agents of the population are evaluated in all 16 possible sensory patterns and gain a reward for correct classification (no reward or penalty for incorrect classifications). The reward values for correct classifications of each class is inversely proportional to their frequency: the reward for PD and ND patterns are 10, and the reward value for correct classification of stationary patterns are 3. However, in the results presented in the next section, all fitness values are normalized to take a maximum value of 100.

\section{Results}
\begin{figure}[htb]
\centering
\includegraphics[scale=0.17]{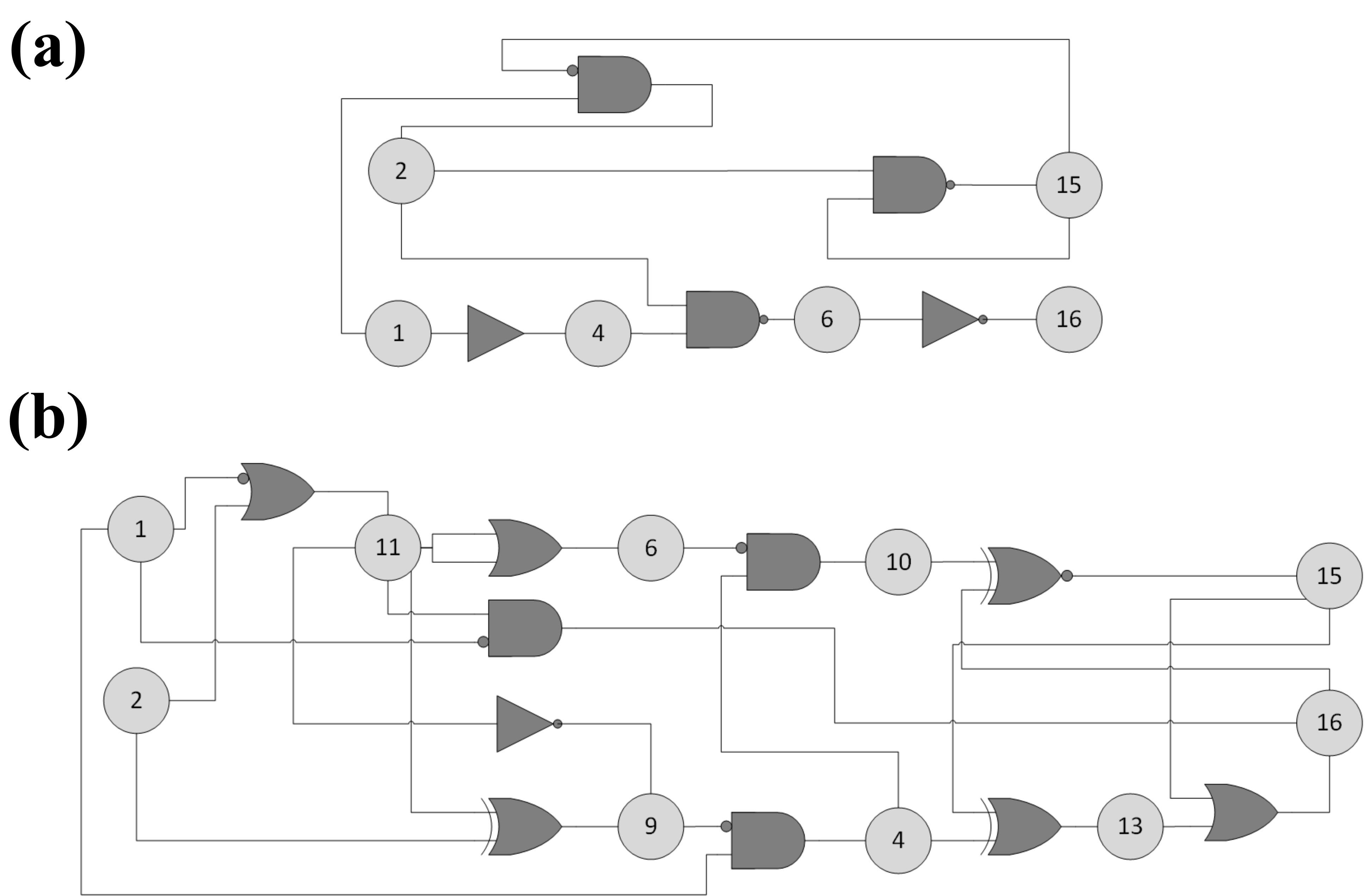}
\caption{Markov brains evolve alternative circuits to encode a motion detection circuit (duplicated logic gates with same inputs and outputs are omitted). a) Example simple evolved motion detection circuit. b) Example complex evolved motion detection circuit. Gate symbols are US Standard.}
\label{fig5}
\end{figure}

After evolving 100 populations for 10,000 generations, we isolated one of the genotypes with the highest score from each population and analyzed its ability to perform the same function as a motion detection circuit. Seventy-five of the one hundred brains evolved a perfect motion detection circuit (correct classification of all 16 patterns); we used those brains for the rest of our analysis. A preliminary analysis of our evolved brains suggested that evolution led to a wide diversity of neuronal circuit architectures. Amongst our population of 75 brains, we found both relatively simple neuronal circuits (Fig.~\ref{fig5}a) and more complex neuronal circuits (Fig.~\ref{fig5}b), suggesting that not only does evolution lead to a large number of different motion detectors, but they also all vary in complexity (defined here as the number of gates composing a circuit).
%Fig. 5

\begin{figure}[htb]
\centering
\includegraphics[scale=0.6]{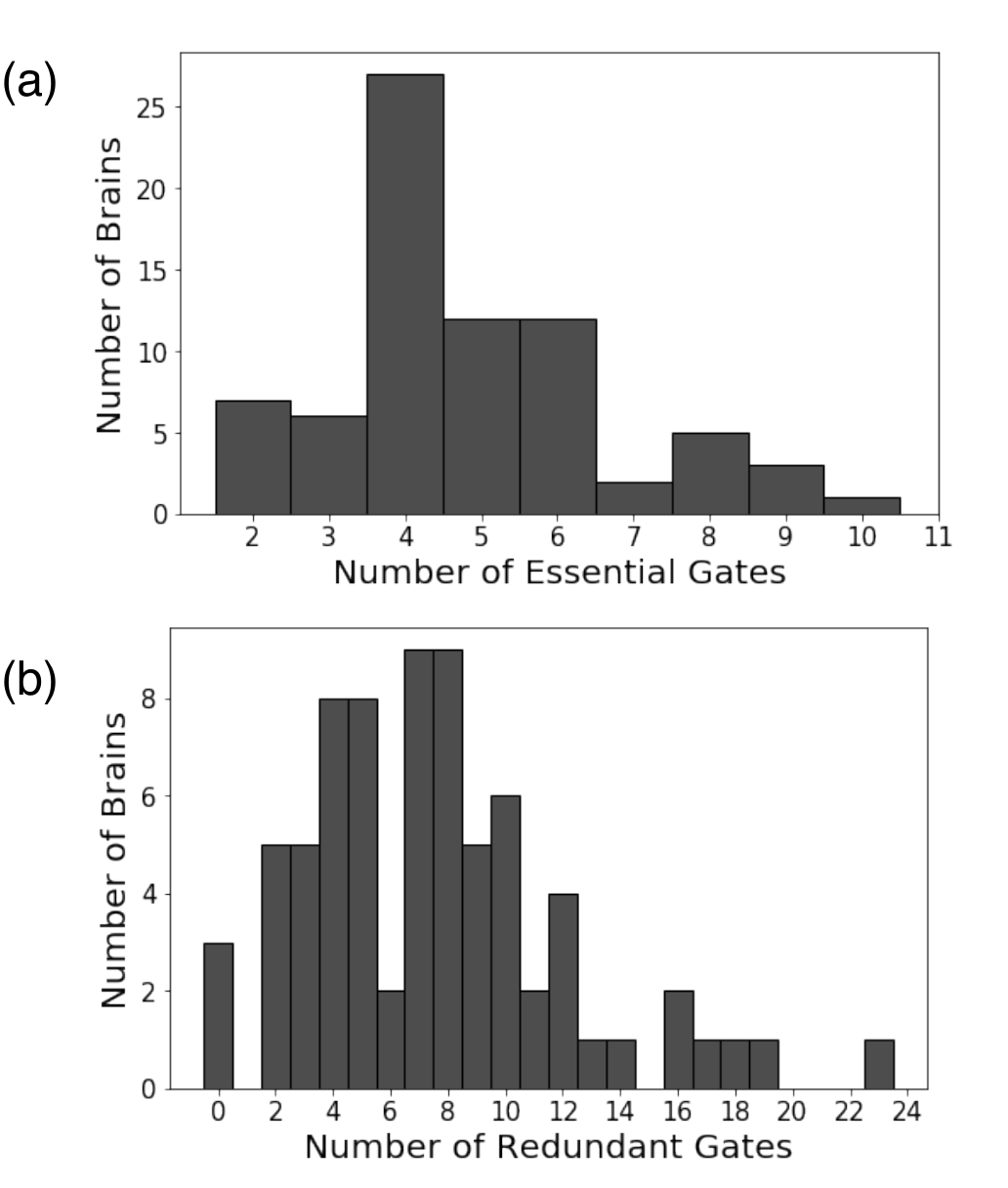}
\caption{Evolved motion detection circuits vary greatly in complexity. a) Histogram of the number of essential gates (i.e., gates that resulted in a fitness loss when removed) for each evolved motion detection circuit. b) Histogram of the number of redundant gates (i.e., gates that resulted in no fitness loss when removed) for each evolved circuit.}
\label{fig6}
\end{figure}

To gain a better understanding of the diversity of neuronal circuits evolved in this study, we performed gate-knockout assays on all 75 brains. We sequentially eliminated each logic gate and re-measured the mutant brain's fitness, thus allowing us to estimate which gates were essential to the motion detection function (if mutant fitness decreased) and which gates were redundant to the motion detection function (if mutant fitness was equal to the ancestral fitness). There was a wide distribution in the number of essential logic gates, ranging from two logic gates to ten logic gates, with a mean of 4.82 gates (Fig.
~\ref{fig6}a). This result supports the idea that there is a wide diversity of possible motion detection circuits available to evolution. We also measured the number of redundant logic gates and found our evolved brains possessed an even greater number of gates that had no apparent contribution to the circuit's function (Fig.~\ref{fig6}b), suggesting that either a large portion of the complexity of these motion detection circuits evolved neutrally, or that selection for redundancy and mutational robustness is involved. 
%Fig. 6

\begin{figure}[htb]
\centering
\includegraphics[scale=0.43]{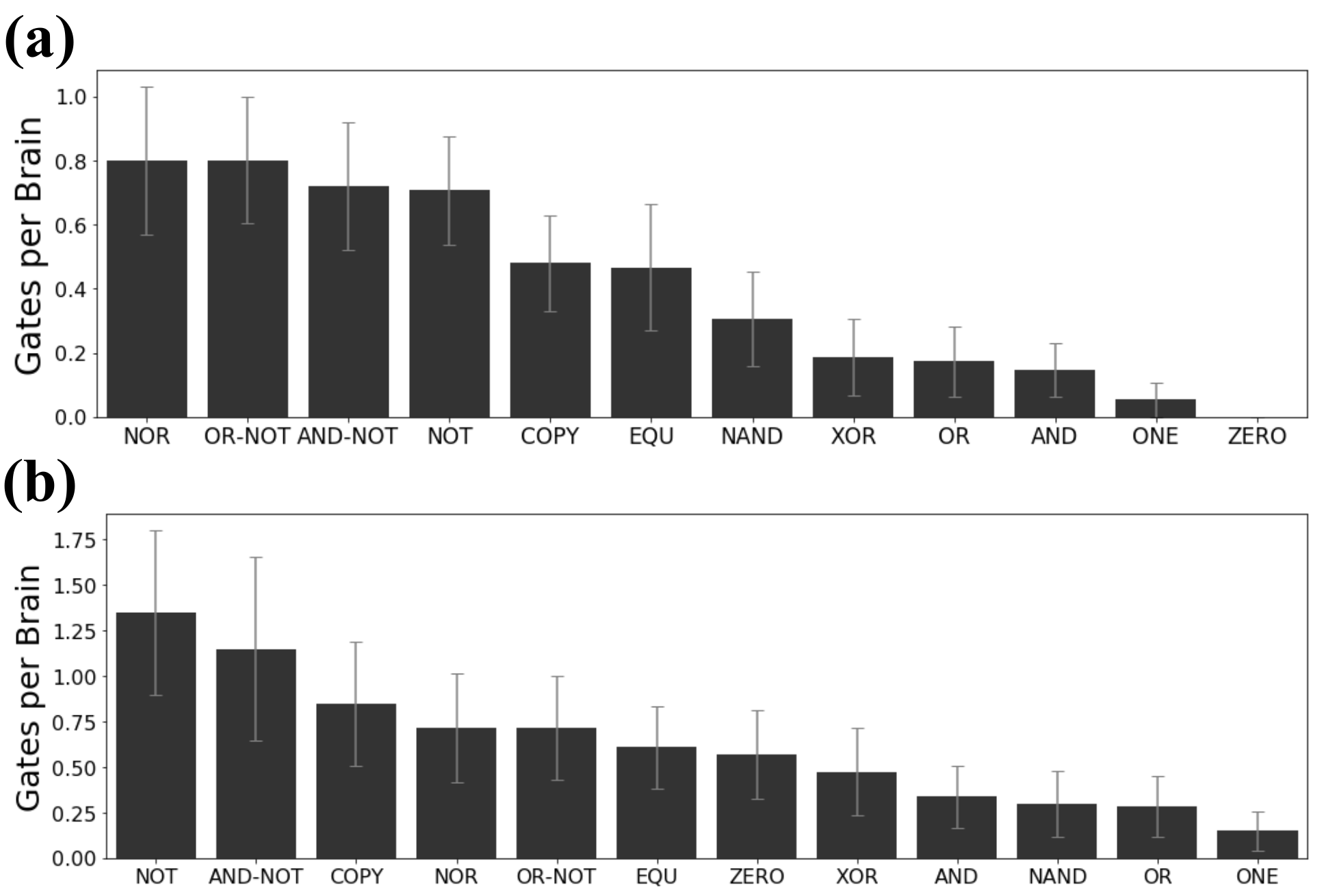}
\caption{Distribution of specific gates used in evolved motion detectors. a) Average number of essential logic gates of each type of logic gate per evolved brain. Error bars represent 95\% confidence intervals. b) Average number of redundant logic gates of each type of logic gate per evolved brain. Error bars represent 95\% confidence intervals.}
\label{fig7}
\end{figure}

We also examined the types of logic gates that were either essential or redundant to each brain by recording the average number each gate was found within each evolved brain. We found surprising similarities in the distribution of the average presence of each logic gate between both essential gates (Fig.~\ref{fig7}a) and redundant gates (Fig.~\ref{fig7}b). The six most-abundant logic gates in both the essential and the redundant gate distribution were NOR, {OR-NOT}, {AND-NOT}, {NOT}, {COPY}, and {EQU}. These results suggest either that evolved motion detection circuits may incorporate whichever gates are most easily-evolved (in the sense that they interact with other gates without fitness trade-offs) or that they may have evolved the same redundant gates as their essential gates in order to encode robustness against mutations.
%Fig. 7

\begin{figure}[htb]
\centering
\includegraphics[scale=0.4]{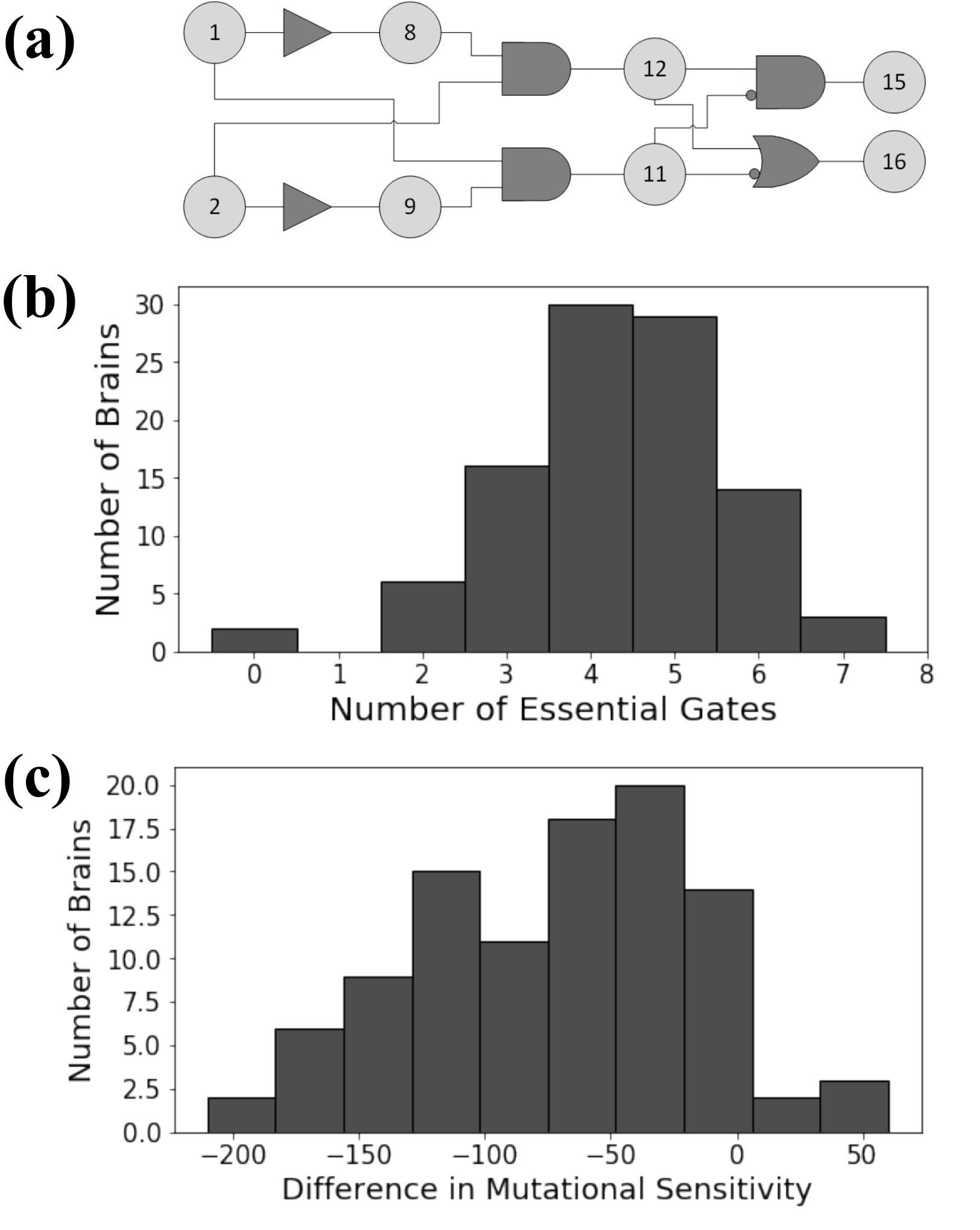}
\caption{Evolution of a simple Reichardt detector leads to greater complexity. a) Diagram of a hand-written Markov Brain encoding a simple Reichardt detector b) Distribution of the number of essential gates for brains evolved from a hand-written ancestor. c) Mutational sensitivity of evolved motion detectors.}
\label{fig8}
\end{figure}

Multiple pieces of evidence suggest that the complexity of our evolved brains did not evolve solely to perform the motion-detection function. Our evolved brains are more complex than required to encode a motion detection circuit (Fig.~\ref{fig6}). 
The large abundance of redundant gates suggests that either these brains are neutrally evolving increased complexity or are evolving mutational robustness due to high mutation rates. The similarities in the distribution of both essential and redundant logic gates suggests either that certain gates arise due to their intrinsic abundance in the fitness landscape, or because they can compensate for mutations to otherwise essential gates. Therefore, to test for the reason behind our evolved brains' complexity, we hand-wrote a simple Reichardt detector with optimal individual fitness (Fig.~\ref{fig8}a), evolved 100 populations under the same protocol as before, and repeated our knockout analysis. If the evolution of complexity was either non-adaptive or due to selection for increased redundancy and robustness, we would expect these simple brains to increase in complexity upon further evolution. However, if the motion detector circuit's evolved complexity is due to difficult-to-break historical contingency, we would expect little change in the brains evolved from hand-written Reichardt detectors. 

The results from the knockout analysis demonstrated that the brains evolved from a hand-written Reichardt detector increased in complexity when evolved further (Fig.~\ref{fig8}b), suggesting that the increased complexity seen in Fig. 6 was not due to historical contingency, but to other evolutionary factors. To test if these evolved brains were shaped by selection for mutational robustness, we measured the mutational sensitivity of each brain by calculating the average fitness loss from removing one logic gate and multiplying this loss by the total number of gates in each brain. Those evolved brains were less mutationally-sensitive (or more mutationally-robust) than their hand-written ancestor (Fig.~\ref{fig8}c), suggesting that the additional gates evolved in order to increase the brain's robustness to mutations. However, we should also note that some brains did evolve a greater mutational sensitivity, suggesting that either robustness was evolved beyond single-step mutations or that there is some role for non-adaptive evolutionary processes in driving circuit architecture.

\section{Discussion}
We tested if a computational model could evolve a wide diversity of neuronal architectures, and studied evolutionary trends in the evolution of these neuronal architectures. We found that selection for motion detection does lead to a wide diversity of neuronal circuits even though each has the same overall function. Most brains are more complex than the standard model for motion detection: the Reichardt detector. Each brain uses many different logic-gate components, although some gates are more common than others. A large portion of the evolved complexity in these brains results from the evolution of redundant gates. We also showed that even hand-written Reichardt detectors increase in complexity when evolved further, suggesting that the large complexity is due to either non-adaptive evolution or selection for functional redundancy. Measurements of the evolved brains' mutational sensitivity suggested they had indeed evolved mutational robustness, illustrating one additional selective pressure beyond basic functionality on the neuronal architecture of motion-detection circuits.

We undertook this study to see if some of the trends detected in the evolution of genetic circuits occurred in the evolution of Markov brains~\citep{tsong2003evolution}.  As found in many other functional systems, including those based on biochemistry~\citep{wagner2011origins} and those based on various digital substrates~\citep{raman2010evolvability,fortuna2017genotype,nitash2017origin}, there is a wide variety of diverse neuronal architectures that can encode a motion-detection circuit that is logically equivalent to that of a Reichardt detector. Our results are in accordance with previous results that showed neuronal circuits with the same functional output could vary between species~\citep{shomrat2011alternative}. These results suggest that a diversity of neuronal architectures may exist for species across life. Our results also suggest that any system with interacting individual components that, when combined, lead to a functioning circuit may possess a diversity of circuits that provide the same function.

While it is perhaps not surprising that our evolved digital brains are different from the default Reichardt detector encoding, we did not expect them to be much more complex. Thus, it is worth discussing how some of our experimental design decisions could have influenced these differences. One likely difference between our evolved brains and real brains is the lack of any fitness cost for larger brains in our model. If each neuron or logic gate was associated with a fitness cost, then one would intuitively expect the evolved brains to be simpler than what we found them to be. On the other hand, neuro-anatomical evidence has suggested that wiring length and connection cost do not appear to be minimized in brains [see also~\citep{HilgetagKaiser2004}]. 

Another difference between digital and biological brains is that we only selected on one trait here. The evolution of neuronal circuits is likely constrained by pleiotropic interactions with other functional circuits, as with genetic systems~\citep{sorrells2015intersecting}. Finally, compared to biological systems, Markov brains evolved under of a very high mutation rate, something that is known to alter the evolution of genetic architecture towards mutational robustness~\citep{wilke2001evolution}. It is likely that Markov brains would have evolved less-complex circuits with a decreased mutation rate, although the magnitude of this effect is not known.

We envision the results we presented here as a first step in establishing Markov brains as a model system to study the potential neuronal architectures evolved by Darwinian natural selection. Some of the limitations discussed above present fruitful avenues for future work that may lead to further insights into the evolutionary potential of biological brains. Although we did not attempt a more-precise classification of our evolved circuits beyond their complexity and their specific logic gates, we see this as a possible endeavor. If the addition of further selection pressures results in the evolution of simpler brains than those evolved here, this task should be achievable. Such studies should lead to a more predictable theory of the diversity of neuronal circuits.

\section{Acknowledgements}
T.L. acknowledges a Michigan State University Distinguished Fellowship, a BEACON fellowship, as well as the Rudolph Hugh and Russell B. DuVall awards, for support. This work was supported in part by Michigan State University through computational resources provided by the Institute for Cyber-Enabled Research, and the National Science Foundation under Grant No. DBI-0939454

%\bibliographystyle{apalike}
%\bibliography{alife2018}

\end{document}